\documentstyle[prl,aps,epsf]{revtex}
\begin{document}

\twocolumn[
\hsize\textwidth\columnwidth\hsize\csname@twocolumnfalse\endcsname

\title{The Glassy Potts Model}
\author{Enzo Marinari}
\address{Dipartimento di Fisica and INFN, 
Universit\`a di Cagliari,\\
via Ospedale 72, 09100 Cagliari (Italy)}
\author{Stefano Mossa}
\address{Dipartimento di Fisica and INFM, Universit\`a di L'Aquila,\\
Localit\`a Coppito, 67010 L'Aquila (Italy)}
\author{Giorgio Parisi}
\address{Dipartimento di Fisica and INFN, 
Universit\`a di Roma {\em La Sapienza},\\
P. A. Moro 2, 00185 Roma (Italy)
}
\date{\today}
\maketitle
\begin{abstract}
We introduce a Potts model with quenched, frustrated disorder, that
enjoys of a gauge symmetry that forbids spontaneous magnetization, and
allows the glassy phase to extend from $T_c$ down to $T=0$.  We study
numerical the $4$ dimensional model with $q=4$ states.
We show the existence of a glassy phase, and we characterize it by
studying the probability distributions of an order parameter, the
binder cumulant and the divergence of the overlap susceptibility.
We show that the dynamical behavior of the system is characterized by
aging. 
\end{abstract}
\twocolumn
\vskip.5pc ] 
\narrowtext


The generalization of the Ising model to a frustrated model containing 
quenched disorder, the Ising spin glass, has provided us with a large 
amount of new physics \cite{MEPAVI}.  Replica Symmetry Breaking has 
been found in the Mean Field theory \cite{PARISI}, and mainly 
numerical simulations strongly hint to its validity in finite 
dimensional disordered Ising spin glasses \cite{MAPARU}.

The need of a generalization of such systems to the Potts models has 
been clear very soon: technical motivations are obvious, while 
physical motivations include the need of describing systems where the 
$Z_{2}$ symmetry of the Ising model is not relevant (real glasses 
being potentially among them \cite{PARGLA}).  The most straightforward 
construction of a Potts spin glass, where the spin variables can be in 
$q$ states and are randomly connected by a positive or a negative 
coupling, has been analyzed in detail \cite{ELDSHE}-\cite{BINDER}, but 
it has the unappealing feature (that we will justify in the following) 
of acquiring a spontaneous magnetization (i.e.  of entering a phase 
with usual ferromagnetic ordering) at low $T$ values.  The glassy
regime is only present in a small $T$ region, making it unpractical 
to be studied numerically and unplausible for a faithfully description 
of real glasses (that do not order at low temperatures).

Here we will define and study what we consider to be the ``naturally 
glassy'' generalization of the Ising spin glasses to the Potts model.  
We will study the finite dimensional version of the model, and we will 
show that these systems do indeed undergo a phase transition that 
leads them to a glassy phase, different from the usual Sherrington 
Kirkpatrick spin glass phase.

We regard here as crucial the exact {\em gauge invariance} that is 
found in the usual Ising spin glasses (both in finite number of 
dimensions and in the mean field approximation).  The Hamiltonian of 
the Ising spin glass is $-\sum \sigma_{i}J_{i,j}\sigma_{j}$, where the 
sum runs over couples of first neighboring sites of the (simple 
hyper-cubic) $d$ dimensional lattice, the $J$ are $\pm 1$ with uniform 
probability (or Gaussian variables), and the spin $\sigma$ take the 
values $\pm 1$.  Let us consider the site $i$ and transform 
$\sigma_{i}\to -\sigma_{i}$.  If at the same time we flip all the $2d$ 
couplings $J_{i,j}$ involving the site $i$ (and one of its first 
neighboring sites $j$) the energy of the system does not change. 
Because of this symmetry the expectation value of the magnetization is 
zero, and a ferromagnetic phase is not allowed.

The generalization to a disordered model of 
\cite{ELDSHE}-\cite{BINDER} is not protected by such a local symmetry, 
and is allowed to magnetize (as it indeed does).  We propose instead a 
generalization to a quenched, frustrated spin model, where the gauge 
invariance is preserved. The Hamiltonian of our model is

\begin{equation}
  H \equiv - \sum_{<i,j>}\delta_{\sigma_{i},\Pi_{i,j}(\sigma_{j})}\ ,
  \label{E-ONE}
\end{equation}
where the sum runs over first neighboring sites on a simple cubic 
hyper-lattice (or over all site couples in the mean field model), the 
spin variables $\sigma$ can take $q$ values ($0$, $1$, $\ldots$, 
$q-1$), and the $\Pi_{i,j}$ are link attached, quenched, random 
permutations of $(0, \ldots,q-1)$ (there are $q!$ of them).  In the 
following we will denote this model by ${\cal M}_{q!}$.  It is clear 
that, when written as a sum of delta functions, the Ising spin glass 
has exactly this form: in this sense this is a very natural 
generalization of the model, where we just increase the number of 
allowed states.  The link $\langle i,j \rangle$ will give a non-zero 
contribution to the action not, as in the usual Potts model, if 
$\sigma_{i}=\sigma_{j}$, but if $\sigma_{i}$ is equal to 
$\Pi_{i,j}(\sigma_{j})$.  In this model the same gauge symmetry we 
have described before protects us against magnetizing: if we transform 
$\sigma_{i}$ from $0$ to $1$ we will interchange in the quenched 
random permutations involving the site $i$ the state $0$ and the state 
$1$.  This feature makes this model (the glassy Potts model) a good 
candidate to the description of the glassy state.

The model we have just defined, ${\cal M}_{q!}$, has a drawback: it is 
very difficult to check if it has reached thermal equilibrium.  The 
$Z_{2}$ symmetry of the Ising model is indeed precious at this end: in 
a spin glass checking the symmetry of the probability distribution of 
the overlap is crucial for establishing thermalization \cite{MAPARU}.  
Our first numerical simulations of the model ${\cal M}_{q!}$ have 
confirmed how difficult it is to establish on firm grounds 
thermalization without being able to count on a slow mode that has to 
exhibit a symmetry.  It is possible to solve this problem at least for 
even values of the number of allowed states, $q$.  One considers a 
permutation $R$ such that $R^{2}=1$ (for example we can change the 
state $2k$ with the state $2k+1$ for $k=0,\ldots\frac{q}{2}-1$), and 
allows in the $\delta$ function only permutations that commute with 
$R$.  We have introduced the model ${\cal M}_{co}$ (where $co$ stands 
for commutative) where the Hamiltonian of equation (\ref{E-ONE}) 
contains only permutation $\Pi$ that commute with the permutation $R = 
(0,1,2,3) \to (2,3,0,1)$, i.e.  $\Pi R = R \Pi$.  This model is 
symmetric under $R$, and invariance under $R$ can be tested in order 
to check if thermal equilibrium has been reached.  In order to do that 
we have defined a modified overlap $\omega$, that is one if two spins 
are in the same state, $-1$ for the couples $(0,2)$, $(1,3)$, $(2,0)$ 
and $(3,1)$, and zero otherwise ($q$ will be the usual overlap, where 
we sum one if two spin are equals and zero otherwise).  Because of the 
symmetry we have introduced by selecting only $R$ commuting 
permutations $\Pi$ the probability distribution $P(\omega)$ is 
symmetric at equilibrium under $\omega \to - \omega$.  The two models 
are expected and turn out to be equivalent, as we will show in the 
following.  When using ${\cal M}_{co}$ it is easy to check 
thermalization, and the coincidence of the results with the ones 
obtained when studying ${\cal M}_{q!}$ shed light on their physical 
meaning.

We have studied ${\cal M}_{q!}$ and ${\cal M}_{co}$ with $q=4$ states
in $4$ spatial dimensions $d$. We have used a normal Monte Carlo
method.  We will present thermalized data in the broken phase for
lattices of volume $L^4$ $=$ $4^{4}$ and $5^{4}$, and data in the warm
phase for a $8^{4}$ lattice. The data on the two smaller lattice
volumes have been obtained from a slow annealing with ten million full
sweeps of the lattice at each temperature point, the one with $L=8$
have used one million sweeps per $T$ point. We have averaged over ten
realizations of the disorder. We have preferred to have long thermal
runs, since thermalization of the samples needs to be completely sure
to make the results reliable, and to keep the number of samples quite
small. The numerical simulations have taken of the order of two months
of medium size workstations.  We only report results for which we are
sure of having reached full thermalization (the main criterion used
being the symmetry of the probability distribution of ${\cal M}_{co}$,
and the request of a good stability in time of the observables).

Working on lattices of linear size $4$ and $5$ we have succeeded to 
get some control over the finite size behavior. A large amount of 
evidence, that we will describe in the following, makes clear the 
existence of a phase transition to a low $T$ glassy phase.

The Binder parameters of the modified overlap 

\begin{equation}
  g_\omega(T) \equiv \frac12 
  \left( 3 -
  \frac
  {\overline{\langle \omega^4\rangle}}
  {\overline{\langle \omega^2\rangle}^2} 
  \right)
\end{equation}
shows the clear signature of a phase transition. For ${\cal M}_{co}$
$g_\omega\simeq 0$ with good accuracy in the warm phase (because of
the symmetry we have implemented). It becomes different from zero at
$T\simeq 1.4$, and grows basically linearly (in our statistical
accuracy) at low $T$. At $T=1.2$ (the lowest $T$ value we are sure we
have thermalized both at $L=4$ and $L=5$) $g_\omega\simeq \frac12$. In
the precision given by our statistical errors (of the average over
disorder), evaluated directly for the Binder parameter by a jack-knife
method, the $L=4$ and $L=5$ results coincide (at $T\simeq 1.2$ the
relative error on the Binder parameter is of he order of ten percent:
for example for $L=5$ we have $g_\omega=0.45 \pm 0.10$). For the model
${\cal M}_{q!}$ $g_\omega$ has the same pattern but it becomes
slightly negative at $T\simeq 1.7$ (but close to zero): at low $T$ it
increases like for the other model. The Binder parameter of the
straightforward overlap does not behave in a interesting way for both
models. From the analysis of the Binder parameters we deduce as a
first guess that $T_c\simeq 1.5$.

\begin{figure}
  \begin{center}
    \leavevmode
    \epsfysize=250pt
    \epsffile{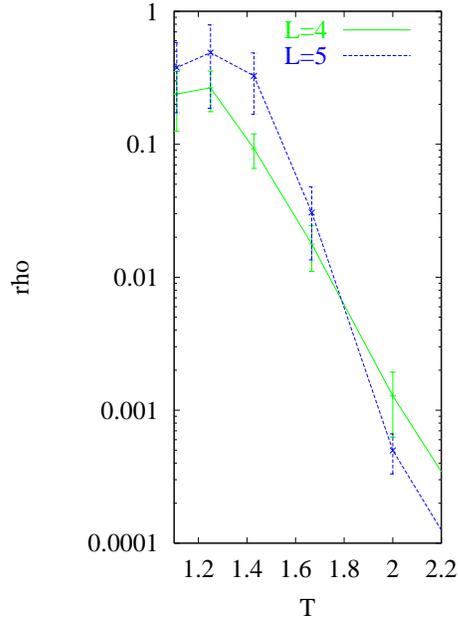}
  \end{center}
  \caption[0]{$\rho(T)$ versus $T$. 
    Solid line for $L=4$ and
    dashed line for $L=5$. 
    Notice the vertical logarithmic scale.}
  \label{F-RSB}
\end{figure}

In figure (\ref{F-RSB}) we plot the Replica Symmetry Breaking
parameter introduced in \cite{MNPRRZ},

\begin{equation}
  \rho_\omega \equiv
  \frac{\overline{\langle \omega^2 \rangle^2}-
        \overline{\langle \omega^2 \rangle}^2}
       {\overline{\langle \omega^4 \rangle}-
        \overline{\langle \omega^2 \rangle}^2}\ ,
\end{equation}
for ${\cal M}_{co}$ (where we have assumed that $\langle\omega\rangle
= 0$. $\rho$ seems to give an even clearer signature of
the phase transition than the Binder parameter (qualifying it without
ambiguities as a Replica Symmetry Breaking transition). In the
infinite volume limit $\rho$ is zero if the overlap distribution is
self-averaging, and becomes non-zero if Broken Replica Symmetry makes
it non-self-averaging. Reference \cite{MNPRRZ} shows that while the
Binder parameter is very effective in detecting symmetry breaking 
accompanied by breaking of spin reversal symmetry, but when spin
reversal symmetry is absent $\rho$ tends to be a better estimator.
Figure (\ref{F-RSB}) shows a sharp transition (notice that the
vertical scale is logarithmic). Again a good estimate for $T_c$ is
$1.5$. The evidence for a phase transition is clear. The fact we can
detect it by using $\rho$ makes clear it is a Replica Symmetry
Breaking transition. Again, the situation in ${\cal M}_{q!}$ is very
similar: there is a change of regime close to $T=1.5$, where $\rho$
does not go to zero with increasing $L$.

\begin{figure}
  \begin{center}
    \leavevmode
    \epsfysize=250pt
    \epsffile{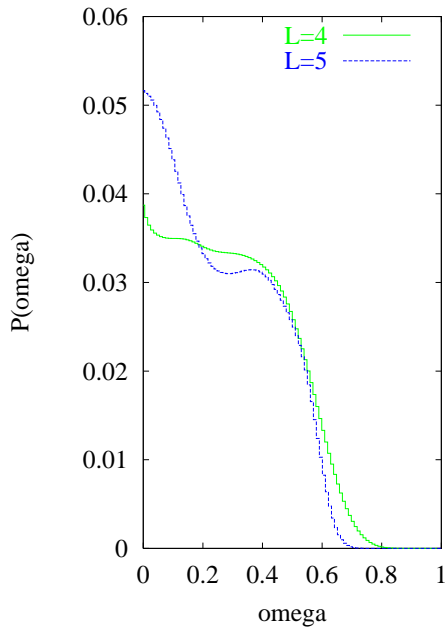}
  \end{center}
  \caption[0]{$P(\omega)$ (symmetrized) versus $\omega$ at
    $T=1.25$. Annealed runs with ten million sweeps per T point.
    Solid line for $L=4$ and
    dashed line for $L=5$. 
  }
  \label{F-PAVE}
\end{figure}

Maybe the most interesting evidence about the behavior of the system
comes from the analysis of the probability distributions of the
overlap averaged over the different disorder samples.  At high $T$
$P(\omega)$ in ${\cal M}_{q!}$ tends to a Gaussian when increasing the
lattice size. We show in figure (\ref{F-PAVE}) $P(\omega)$ at $T=1.25$
for $L=4$ and $5$. There is a clear non-trivial behavior (notice that
the support in $\omega$ is very extended, i.e. $P(\omega)$ is very
flat). On larger lattices a peak in $\omega\simeq 0$ is emerging,
separated by a flat minimum from a second peak (the minimum becomes
sharper at lower $T$ values, but there we know we did not reach full
thermal equilibrium and we cannot safely attach a firm significance to
the data): this behavior is reminiscent of the one of REM models, and
fits with what we expect for a glassy state \cite{PARGLA}. The
important point here is that we have been able to thermalize the $L=5$
system even in the cold phase ($T=1.25$ is the lowest temperature
value where we are sure about an adequate thermalization).

\begin{figure}
  \begin{center}
    \leavevmode
    \epsfysize=250pt
    \epsffile{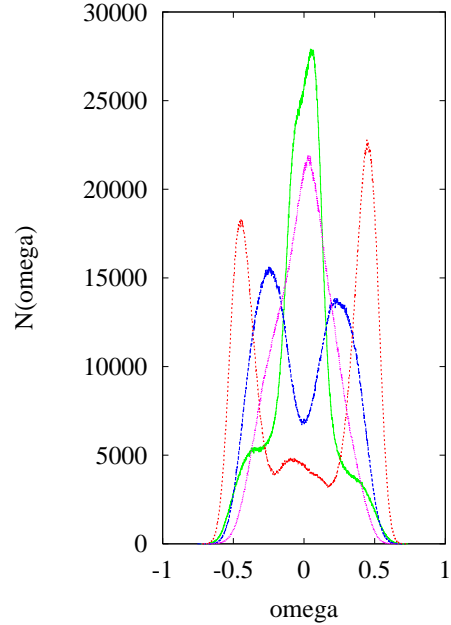}
  \end{center}
  \caption[0]{
    $N_J(\omega)$ (the non-normalized $P_J(\omega)$) for
    four typical samples, versus $\omega$ at
    $T=1.25$. 
    Annealed runs with ten million sweeps per T point.
  }
  \label{F-PFOUR}
\end{figure}

It is useful (and necessary) to look at the individual, single sample
probability distributions $P_J(\omega)$ in order to qualify the
behavior of the individual systems. In figure (\ref{F-PFOUR}) we plot
$N_J(\omega)$ (the non-normalized $P_J(\omega)$) for four typical
samples, versus $\omega$ at $T=1.25$. The level of the asymmetry of
the histograms is a measure of our statistical error (and the fact the
functions look symmetric a sign of a good thermalization). The
$N_J(\omega)$ are non-trivial: some samples have double peaks, some
have their support close to $\omega=0$, some have support at zero
overlap and peaks at finite overlap. On a qualitative level we remark
that system looks {\em harder} than the usual spin glass: the dynamics
is more jumpy, visiting in a quite discontinuous manner different
parts of the phase space (that is why checking thermalization has been
difficult and crucial). Our evidence seems to suggest that the free
energy phase space is {golf course} like: deep minima do not have
large basins of attraction.

The probability distribution for the modified overlap $\omega$ in the
model ${\cal M}_{q!}$ is similar to the one we have shown. The
detailed shape is not exactly the same, but it also becomes
non-trivial at $T=1.25$. On the $L=5$ lattice a two peak structure
starts to emerge from $T=1.25$ down (one in $\omega\simeq 0$ and one
at a finite $\omega$ value). The non-modified overlap distributions for
both models enjoy the same main features (even if they are
non-symmetric, and, as we have already discussed, the thermalization
is better checked by using the symmetry of $P(\omega)$ in ${\cal
M}_{co}$): it tends to a Gaussian when increasing $L$ at $T=1/0.6$, it
is non-trivial at $T=1/0.7$, and it develops a two peak structure at
lower $T$ values.

We have used the $T$ data in the warm phase from the large, $L=8$
lattice to fit the divergence of the overlap susceptibility
$\chi_\omega \equiv V\langle \omega^2\rangle$, that  $T\to T_c^+$
behaves as $(T-T_c)^{-\gamma}$. From
our data we can only give a preliminary estimate, that puts $T_c$
among $1.4$ and $1.5$ (compatible with the value we have deduced from
the direct analysis of $P(\omega)$) and the exponent $\gamma$ among
$1.3$ and $1.5$. This value is different from the one quoted for the
Ising $4D$ spin glass, $\gamma = 2.10$ \cite{PARIRU}.

\begin{figure}
  \begin{center}
    \leavevmode
    \epsfysize=250pt
    \epsffile{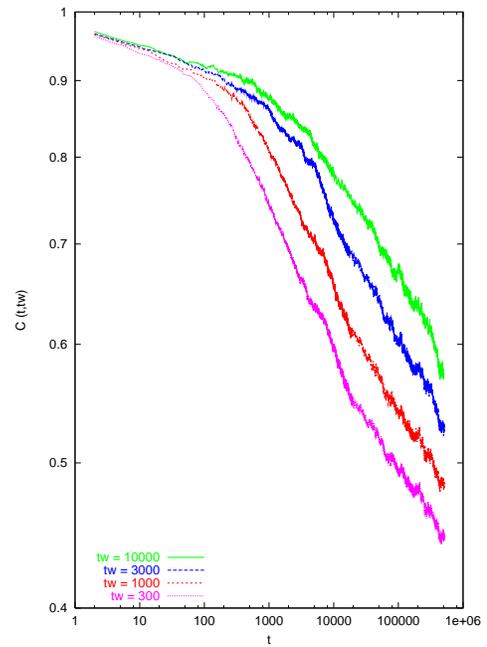}
  \end{center}
  \caption[0]{
    Dynamical correlation function $C(t,t_w)$ versus $t$ for different
    values of $t_w$.
  }
  \label{F-AGING}
\end{figure}

The dynamical behavior of the system shows all the typical features of
the complex dynamics. In figure (\ref{F-AGING}) we show the aging
behavior: the spin-spin correlation function $C(t,t_w)$, depending on
the waiting time $t_w$ and on the time $t$, versus $t$ for different
waiting times $t_w$. The rate of the time decay depends on $t_w$.

Let us notice at last that the energies of the two models are very
similar: on the $L=5$ lattice they are equal in our statistical
precision, while on the $L=8$ lattice they are of the order of one per
one thousand. The two models seem to present the same kind of critical
(and off-critical) behavior, and our best guess is that they do belong
to the same universality class.

We have introduced a disordered generalization of the Potts model that
we regard as a very hopefully candidate to describe the glassy
state. Our numerical simulations of the 4 state, 4 dimensional model
show clearly the existence of a glassy phase, and they stress large
differences with the usual Sherrington Kirkpatrick spin glass phase,
exhibiting a more discontinuous behavior, reminiscent of the Random
Energy Model one step Replica Symmetry Breaking \cite{MEPAVI}.

We thank Renate Loll for the gift of a nifty script computing commuting 
permutations.


\end{document}